\documentclass{article}

\usepackage{arxiv}

\usepackage[utf8]{inputenc} 
\usepackage[T1]{fontenc}    
\usepackage{hyperref}       
\usepackage{url}            
\usepackage{booktabs}       
\usepackage{amsfonts}       
\usepackage{nicefrac}       
\usepackage{microtype}      
\usepackage{lipsum}
\usepackage{graphicx}
\graphicspath{ {./images/} }
\usepackage{amsmath, amssymb}

\usepackage{graphicx}%
\usepackage{multirow}%
\usepackage{amsmath,amssymb,amsfonts}%
\usepackage{amsthm}%
\usepackage{mathrsfs}%
\usepackage[title]{appendix}%
\usepackage{xcolor}%
\usepackage{textcomp}%
\usepackage{manyfoot}%
\usepackage{booktabs}%
\usepackage{algorithm}%
\usepackage{algorithmicx}%
\usepackage{algpseudocode}%
\usepackage{listings}%
\usepackage{natbib}
\usepackage{todonotes}
\usepackage{verbatim}

\newtheorem{proposition}{Proposition}%

\title{Spatial Gaussian fields for complex areas with application to marine megafauna conservation}

\author{Martina Le-Bert Heyl\\
 Marine Science Department, KAUST, Saudi Arabia \\
  \texttt{martina.lebertheyl@kaust.edu.sa, martinaleberth@gmail.com} \\
   \And
Janet van Niekerk \\
  Statistics Program, KAUST, Saudi Arabia\\
  Department of Statistics, University of Pretoria,
  South Africa\\
  \texttt{janet.vanniekerk@kaust.edu.sa, janet.vanniekerk@up.ac.za} \\
  \And
 H{\aa}vard Rue \\
  Statistics Program, KAUST, Saudi Arabia\\
  \texttt{haavard.rue@kaust.edu.sa} \\
}

\begin{document}
\maketitle
\begin{abstract}
Spatial Gaussian fields (SGFs) are widely employed in modeling the distributions of marine megafauna, yet they traditionally rely on assumptions of isotropy and stationarity, conditions that often prove unrealistic in complex ecological environments featuring coastlines, islands, and depth gradients acting as partial movement barriers. Existing spatial models typically treat these barriers as either fully impermeable, completely blocking species movement and dispersal, or entirely absent, which inadequately represents most real-world scenarios.
To address this limitation, we introduce the Transparent Barrier Model, an extension of spatial Gaussian fields that explicitly incorporates barriers with varying levels of permeability. The model assigns spatially varying range parameters to distinct barrier regions, allowing ecological and geographical knowledge about barrier permeability to directly inform model specifications. This approach maintains computational efficiency by utilizing the integrated nested Laplace approximation (INLA) framework combined with stochastic partial differential equations (SPDEs), ensuring feasible application even in large, complex spatial domains.
We demonstrate the practical utility and flexibility of the Transparent Barrier Model through its application to dugong (Dugong dugon) distribution data from the Red Sea. This area is characterized by a heterogeneous spatial structure, influenced by coastlines, islands, and varying bathymetric depths. Our results indicate that the model effectively captures spatial dependencies and gradients of barrier permeability, significantly enhancing spatial inference accuracy. This methodological advancement offers ecologists and conservation planners a powerful and realistic tool for modeling species distributions, facilitating more informed decisions in marine conservation and spatial management across diverse ecological landscapes.
\end{abstract}

\keywords{Permeable Barriers, Species distribution models, SGF, INLA, SPDE, Marine Megafauna, Dugong, Red Sea, Saudi Arabia}

\section{Introduction}\label{sec1}
Marine megafauna, such as whales, sharks, and dugongs, play pivotal roles in marine ecosystems. Often regarded as keystone species, they influence the abundance and distribution of other marine organisms \cite{estes_trophic_2011, estes_megafaunal_2016, pimiento_functional_2020, lambert_energyscapes_2025}. Dugongs, for instance, significantly contribute to the health of seagrass ecosystems by grazing on seagrass beds. This grazing activity prevents overgrowth, promotes habitat complexity, and supports diverse marine life \cite{burkholder_patterns_2013, rasul_grazing_2024, shawky_dugong_2024}. Consequently, conservation efforts targeting marine megafauna are vital to maintain marine ecosystem resilience.

Species distribution models (SDMs) provide valuable insights into species distributions, habitat preferences, and ecological interactions. Traditional SDMs predict species occurrences using environmental variables, such as water temperature, depth, and ocean currents, employing methods like MaxEnt, Random Forest, and Generalized Additive Models (GAMs) (e.g. \cite{roberts_habitat-based_2016, wang_mapping_2020, reiss_species_2011, darr_detecting_2014}). However, the predictive accuracy of these models heavily depends on comprehensive environmental data availability limiting traditional SDMs \cite{mcclellan_understanding_2014-1, gonzalez-irusta_comparing_2015}. When comprehensive covariate information is lacking, modeling spatial random effects becomes crucial to account for unexplained spatial dependencies \cite{legendre_spatial_1989, guisan_predicting_2005, pollock_understanding_2014}. 

Spatial Gaussian fields (SGFs) are widely used in spatial and spatio-temporal modeling, particularly when addressing residual spatial structures resulting from unmeasured covariates, spatial aggregation, or spatial noise \cite{diggle_model-based_1998, gelfand_change_2001, rue_gaussian_2005, cameletti_spatio-temporal_2013}. Among SGFs, the Matérn covariance model \cite{whittle_stationary_1954} is widely adopted due to its flexibility and computational efficiency, further enhanced by recent methodological advancements such as the integrated nested Laplace approximation (INLA) \cite{rue_approximate_2009} and the stochastic partial differential equation (SPDE) \cite{lindgren_explicit_2011} approach, facilitating efficient Bayesian inference \cite{blangiardo_spatial_2015, krainski_advanced_2018, rue_bayesian_2017, bakka_spatial_2018}.

Despite these advantages, standard SGFs typically rely on assumptions of stationarity and isotropy, meaning the spatial autocorrelation structure remains invariant under spatial translation and rotation. These assumptions often become unrealistic in complex geographical environments characterized by physical barriers, such as islands or coastlines \cite{wood_soap_2008, scott-hayward_complex_2014, bakka_non-stationary_2019}. In these contexts, spatial dependence should not solely depend on Euclidean distance, as barriers significantly influence spatial connectivity \cite{whitaker_geographic_2003, nislow_variation_2011, letcher_population_2007}.

The Barrier model proposed by \cite{bakka_non-stationary_2019} addresses this limitation by extending the Matérn framework to non-stationary settings \cite{martinez-minaya_dealing_2019, chaudhuri_modeling_2023}. Instead of relying on the shortest Euclidean distance to determine spatial dependence, the Barrier model accounts for all potential connections between two locations. When physical barriers are present, the paths that cross over the barrier are removed, possibly weakening the overall dependency between the two locations \cite{bakka_non-stationary_2019} as is appropriate.

While the Barrier model is limited to physically impermeable barriers, many real-world scenarios involve barriers with varying permeability \cite{pepino_fish_2012, cozzi_comparison_2013, altermatt_diversity_2013,  altermatt_diversity_2013, pepino_thermal_2024}. For instance, islands may act as impermeable barriers for specific marine species, whereas sand patches with tidal water coverage may allow partial movement of others \cite{grantham_dispersal_2003, birk_three_2012}. \cite{li2023design} proposed a multi-barrier model using the INLA-SPDE framework, but the implementation of the model and a definition of transparency is omitted.

In Section \ref{sec:mot} we present the motivating example of species distribution modeling of dugongs on the coast of the Red Sea in Saudi Arabia. Due to specific dugong behavior and the complicated coastal area we have various areas in the spatial domain with different properties and permeability such as sandy areas, deep sea, canals and so one.\\  

To address this naturally occurring phenomena, we propose the Transparent Barrier model, an innovative adaptation of the Barrier model framework that accounts for barriers with varying permeability, handling both fully impermeable and partially permeable barriers within the same framework. We refer to the latter as transparent or permeable barriers.
 This model retains the computational efficiency of stationary models, since the size of the graph for the stationary, barrier and transparent barrier model are the same, while addressing the complexity of spatial structures influenced by barriers of varying nature, making it a practical and versatile tool for spatial modeling in marine science and other domains.\\

 We propose the Transparent Barrier model (TBM) in Section \ref{sec:meth} and present an intuitive method to incorporate user-knowledge of preconceived transparency of the barrier. We also define the log-Gaussian Cox process (LGCP) accounting for permeable barriers to model a point process. We conduct a simulation study with the aim of investigating certain properties of the TBM and illustrate that the TBM has the Matérn model and the Barrier model as boundary cases, in Section \ref{sec:sim}. We apply the method to obtain focus areas of conservation based on species distribution modeling of dugongs, along the northern coast of the Red Sea in the midst of immense touristic development, based on incidental sightings in Section \ref{sec:app}. The paper is concluded by a discussion in Section \ref{sec:disc}.

\section{Motivating example}\label{sec:mot}
Dugongs (\textit{Dugong dugon}) are globally classified as vulnerable to extinction due to various anthropogenic threats \cite{marsh_dugong_2009}. Long-term and intensive human activities have dramatically reduced dugong populations, leaving fragmented and isolated groups at risk of local extinction \cite{marsh_ecology_2011, marsh_sirenian_2012, dsouza_long-term_2013, dsouza_seagrass_2015, seal_spatial_2024, wang_spatial_2025}. Currently, significant dugong populations are restricted to select areas, primarily around Australia, New Caledonia, Papua New Guinea, the Arabian Gulf, and the Red Sea \cite{preen_distribution_2004, marsh_dugong_2009, shawky_dugong_2024, wang_spatial_2025}. Consequently, global dugong conservation efforts face substantial biological, ecological, and governance-related challenges, characteristic of large, wide-ranging marine herbivores. These challenges are particularly acute along the rapidly developing northern coast of the Saudi Arabian Red Sea, where human activity is rapidly growing \cite{khamis_identifying_2022}. 

Dugongs inhabit shallow coastal waters where their primary food source, seagrass, is found. Seagrass meadows, which are most abundant in intertidal and subtidal zones, typically grow at depths shallower than 10 meters, although some species may extend slightly deeper depending on water clarity and light availability \cite{chilvers_diving_2004, sheppard_dugong_2010}. Dugong movements closely follow tidal cycles, foraging in intertidal zones during high tides and retreating to deeper waters during low tides \cite{sheppard_effects_2009}. While shallow seagrass habitats dominate their diving behavior with most of their activity occurring near the surface \cite{chilvers_diving_2004, sheppard_movement_2006}, dugong movements exhibit considerable complexity influenced by various ecological and anthropogenic factors. For example, dugongs alter their behavioral patterns, such as reducing resting periods and frequently switching between foraging and traveling, in response to predation risk from tiger sharks, particularly in shallow habitats perceived as high-risk \cite{wirsing_behavioural_2012}. Additionally, they adapt their spatial usage, favoring deeper, safer waters despite the greater foraging benefits in shallow zones when predator presence increases \cite{wirsing_living_2007, heithaus_towards_2009}. Human activities further complicate dugong habitat use, with ongoing threats such as bycatch and illegal hunting significantly influencing their distribution patterns, leading to notable occupancy declines and habitat shifts over time \cite{dsouza_long-term_2013, shawky_et_al_assessing_2024, wang_spatial_2025}. Moreover, spatial habitat suitability for dugongs is influenced by a range of environmental factors beyond seagrass presence, including bathymetry, depth gradients, and proximity to shore \cite{seal_spatial_2024}.

While advanced technologies such as aerial surveys and unmanned aerial vehicles are ideal for monitoring dugong presence and abundance \cite{maire_detection_2013, pollock_estimating_2006, hodgson_unmanned_2013}, these resources are often inaccessible in regions experiencing rapid coastal development and urgent conservation needs. Consequently, alternative approaches, such as identifying seagrass feeding trails, are valuable initial methods for informing conservation strategies \cite{khamis_identifying_2022}. However, given the complexity of dugong behavior and the numerous factors influencing their spatial distribution, methodologies should explicitly account for spatial dependencies product of underlying patterns that cannot be explained by measured covariates \cite{legendre_spatial_1989, guisan_predicting_2005, pollock_understanding_2014}. This consideration is especially pertinent at local scales where data are limited \cite{dsouza_long-term_2013}.

\begin{figure}[h!]
    \includegraphics[width = 14cm]{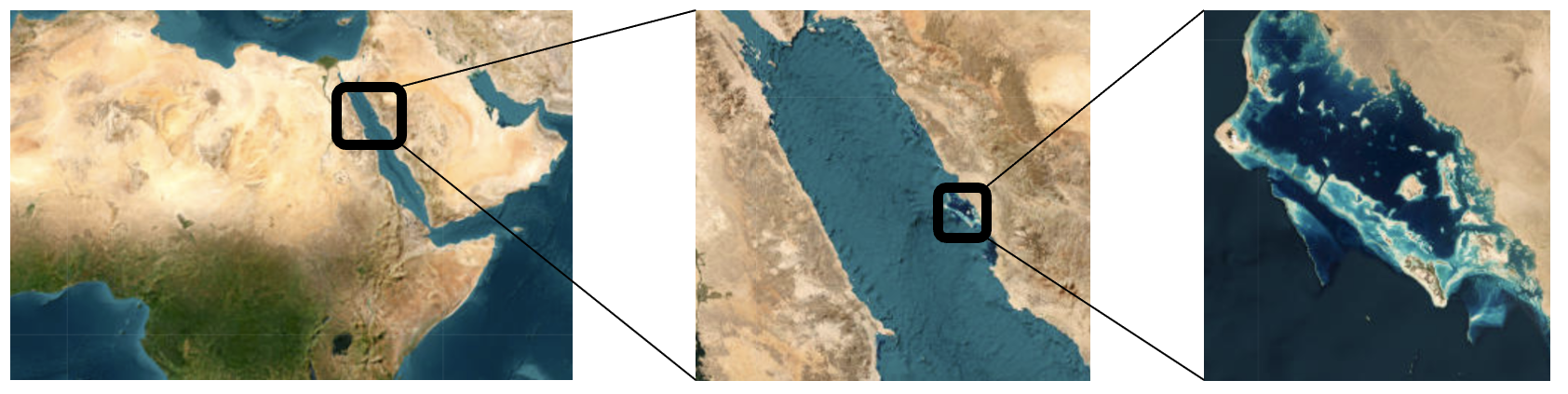}
\label{fig:mot}
\caption{Study area in the Red Sea for Dugong conservation}
\end{figure}

Explicit spatial modeling of species distribution near physical barriers such as coastlines or islands introduces additional challenges. Our study area for the application in Section \ref{sec:app} is on the coast of the Red Sea as illustrated in Figure \ref{fig:mot}, and the complex environment is clear. Standard modeling methods may erroneously smooth spatial autocorrelation across impermeable barriers like sand islands, artificially extending dependencies from marine environments onto land \cite{bakka_non-stationary_2019}. 
Moreover, we need to address whether spatial autocorrelation should be equally smoothed over areas where dugongs can move freely and areas where permeable barriers are present, that is areas where dugongs are less likely to be found, such as deep areas, due to their diving behavior \cite{chilvers_diving_2004, sheppard_movement_2006}


By developing a spatial model incorporating random effects and varying permeability of physical barriers, our study aims to enhance understanding of dugong distribution and inform conservation efforts. This modeling framework not only addresses the specific challenges encountered in the Red Sea but also provides a generalizable methodology for other marine and ecological studies in complex spatial environments. 


\section{Methodology}\label{sec:meth}

\subsection{Preliminaries}
Suppose that we observe data $y (s)\in \mathbb{R}$ at locations $s\in \Omega\subset\mathbb{R}^2$. Based on some observed covariates $\pmb X$, we can formulate the following spatial model,
\begin{equation*}
 Y| \beta, u \sim g( u ,\cdot),\quad u  ( s) =  \beta^\top \pmb X +  u( s),
\end{equation*}
such that $ u( s)$ is a latent spatial field, used to generate the data through the data-generating model $g(\cdot)$ (not necessarily Gaussian) and the spatially varying mean $ u  ( s)$. From a Bayesian viewpoint we can formulate a prior for $ u( s)$ to enforce smoothness and some other properties so that the spatial model becomes
\begin{equation*}
 Y| \beta, u, \theta \sim g( u , \theta),\quad  u  ( s) =  \beta^\top \pmb X +  u( s), \quad 
 u( s)| \omega \sim h( \omega),\label{eq:model1}
\end{equation*}
while also assigning appropriate priors for $ \beta,  \theta, \omega$. \\
Note that if we assume a Gaussian prior for the latent spatial field, i.e. $ u( s)| \omega \sim N\left( u _{ u} ,  \pmb\Sigma_{ u} ( \omega)\right)$, then \eqref{eq:model1} is a latent Gaussian model which enables efficient Bayesian inference and $ u( s)$ is termed a spatial Gaussian field (SGF).\\ \\
A common SGF is defined through the 
Matérn covariance function \cite{whittle_stationary_1954, stein_interpolation_1999, gelfand_handbook_2010}, defined as a function of the Euclidean distance $\|s_i - s_j\|$ between locations $s_i$ and $s_j$, as given by:

\begin{equation}
\mathrm{Cov}(u(s_i), u(s_j)) = \sigma_{u}^{2}\frac{2^{1-\nu}}{\Gamma(\nu)}\left(\kappa\|s_i - s_j\|\right)^{\nu}K_{\nu}\left(\kappa\|s_i - s_j\|\right),\label{eq:matern}
\end{equation}
where $\sigma_{u}^{2}$ is the marginal variance, $K_{\nu}$ is the modified Bessel function of the second kind, and $\nu > 0$, $\kappa > 0$ control smoothness and scale, respectively.

By setting $\nu = 1$ and adopting the empirical definition $\kappa = \sqrt{8\nu}/r$, we obtain:
\begin{equation}
\mathrm{Cov}(u(s_i), u(s_j)) = \sigma_{u}^{2}\left(\frac{\sqrt{8}}{r}\|s_i - s_j\|\right) K_{1}\left(\frac{\sqrt{8}}{r}\|s_i - s_j\|\right),
\end{equation}
where the range parameter $r = \sqrt{8\nu}/\kappa$ indicates the distance at which the spatial correlation between two points approximates $0.13$, thus providing a reparameterization of the Matérn covariance function with interpretable parameters, $\sigma_u$ and $r$.

The Matérn SGF can be efficiently inferred based on the approximate Bayesian framework termed the integrated nested Laplace approximation (INLA) methodology \cite{rue_approximate_2009, van2023new} due to the explicit link between the Matérn SGF and the weak solution of a specific stochastic partial differential equation (SPDE) as shown by \cite{lindgren_explicit_2011}. \\ \\
Suppose that $ u( s)| \sigma_u, r \sim N\left( 0,  \pmb\Sigma_{ u} (\sigma_u, r)\right)$ where the elements of $ \pmb\Sigma_{ u} (\sigma_u, r)$ is defined as in \eqref{eq:matern}, then $ u( s)$ is the weak solution of the following SPDE,

\begin{equation}
 u(s)-\nabla \cdot \frac{r^2}{8} \nabla u(s)=r \sqrt{\frac{\pi}{2}} \sigma_{u} \mathcal{W}(s), \quad\text { for } s \in \Omega,
\end{equation}
where $\mathcal{W}(s)$ is a Gaussian white noise process over $ \Omega$.
This SPDE can be solved using the Finite Element Method (FEM) and the solution leads to a Gaussian field $u(s)$ with sparse precision matrix $ Q_u$ constructed from the elements of the FEM, that is approximately the inverse of the Matérn covariance matrix $ \Sigma_{ u} (\sigma^2_u, r)$. More details are available in \cite{lindgren_explicit_2011}.\\

This INLA-SPDE framework, mitigates the computational challenges associated with large covariance matrices by utilizing sparse precision matrices. Nonetheless, this SPDE framework can be used in other Bayesian inference architectures and is not specific to INLA, although INLA is optimally tailored for sparse matrix computations. The Matérn model can deployed not only in Gaussian likelihood settings, but also in complex scenarios such as point process models (see Section \ref{sec:app}) and marked point pattern models \cite{illian_toolbox_2012}.\\ 

Since the Matérn covariance depends on the Euclidean distance, the SGF is assumed stationary and isotropic, meaning the spatial process is invariant to translation and rotation. In the presence of physical barriers like coastlines, the Matérn model fails by spuriously modeling correlation through the physical barrier. To this end, the barrier model was introduced by \cite{bakka_non-stationary_2019}.\\ \\

The Barrier model extends the Matérn SGF to a non-stationary Matérn SGF while incorporating the model in the INLA-SPDE framework for efficient inference. The premise of this model is to formulate distinct Matérn SGFs to barrier and non-barrier areas. In the Barrier model, separate Matérn fields are specified for normal and barrier areas. Thus, $u (s)$ is now defined by two SPDES as follows:

\begin{align}
& u (s)-\nabla \cdot \frac{r_n^2}{8} \nabla u (s)=r_n \sqrt{\frac{\pi}{2}} \sigma_{u } \mathcal{W}(s), \quad\text { for } s \in \Omega_n \nonumber \\
& u (s)-\nabla \cdot \frac{r_b^2}{8} \nabla u (s)=r_b \sqrt{\frac{\pi}{2}} \sigma_{u } \mathcal{W}(s), \quad\text { for } s \in \Omega_b, \label{eq2}
\end{align}
where $\Omega_n$ and $\Omega_b$ are normal and barrier areas, respectively, and their union covers the entire study area $\Omega$. Parameters $r_n$ and $r_b$ denote the spatial range in normal and barrier areas, with $r_b = r_n/h$ (usually with $h$ large, e.g., $h=10$) to reduce correlation across barriers. Here, $\nabla = \left(\frac{\partial}{\partial x}, \frac{\partial}{\partial y}\right)$ and again $\mathcal{W}(s)$ denotes Gaussian white noise on $\Omega$.

Note that now the range and standard deviation are spatially varying parameters, i.e. $r=r(s), \sigma_u = \sigma_u(s)$. This approach naturally results in increased prior uncertainty in regions such as narrow inlets, where spatial constraints influence movement patterns. This aligns with real-world spatial processes where areas with restricted movement exhibit greater variation in occupancy, as they are either highly concentrated or nearly unoccupied. This interpretation emerges naturally from the SPDE framework, where the resultant field represents the expected location of a randomly moving point.\\ \\
In the next section, we propose a non-stationary Matérn-type SGF model for multiple exhaustive partitions of the study area $\Omega$.

\subsection{Transparent Barrier Model}\label{sec:tbm}

Suppose the study domain $\Omega$, is partitioned into $k$ subdomains $\Omega_q$, with $q \in \left\{1,...,k \right\}$, and  $\bigcup_{q=1}^{k} \Omega_q = \Omega$. With $q=1$, $\Omega_1 = \Omega_n$ the subdomain representing the normal area, and with $q=2,...,k$, $\bigcup_{q=2}^{k} \Omega_q = \Omega_b$ the union of the subdomains representing the various barrier areas. A common marginal variance is assumed, but the range $r_q$ can be particular to each $\Omega_q$.\\ \\

Then, we define the SGF $u(s)$ by the system of SPDEs as follows:

\begin{align}
& u (s)-\nabla \cdot \frac{r_1^2}{8} \nabla u (s)=r_1 \sqrt{\frac{\pi}{2}} \sigma_{u } \mathcal{W}(s), \quad\text { for } s \in \Omega_n \nonumber \\  
& u (s)-\nabla \cdot \frac{r_{2}^2}{8} \nabla u (s)=r_{2} \sqrt{\frac{\pi}{2}} \sigma_{u } \mathcal{W}(s), \quad\text { for } s \in \Omega_{2} \nonumber \\
& u (s)-\nabla \cdot \frac{r_{3}^2}{8} \nabla u (s)=r_{3} \sqrt{\frac{\pi}{2}} \sigma_{u } \mathcal{W}(s), \quad\text { for } s \in \Omega_{3} \nonumber \\
& \vdots \quad\quad\quad\quad\quad\quad\quad\quad\quad\quad \vdots \nonumber \\ 
& u (s)-\nabla \cdot \frac{r_{k}^2}{8} \nabla u (s)=r_{k} \sqrt{\frac{\pi}{2}} \sigma_{u } \mathcal{W}(s), \quad\text { for } s \in \Omega_{k}. \label{eq3}
\end{align}

Similarly to the stationary Matérn SGF we use the FEM to solve this system of SPDES, requiring a Delaunay triangulation mesh that discretizes the spatial domain into triangles. Recall that the FEM is used to construct a sparse precision matrix for approximation of the SGF $u(s)$, that enables efficient inference for this model using sparse-specific framework like INLA. At each mesh node, we define a linear finite element basis function $\psi_i(s)$, equal to $1$ at node $i$ and $0$ elsewhere. The approximation of the Gaussian field $\tilde{u}(s)$ is then given by

\begin{equation}
\tilde{u}(s) = \sum_{i=1}^{n}u_i\psi_i(s),
\end{equation}
where $u_i$ are Gaussian-distributed coefficients with precision matrix $\mathbf{Q}$. Rewriting the SPDE gives

\begin{equation}
\left[1 - \nabla \frac{r(s)^2}{8}\nabla\right]u(s) = r(s)\sqrt{\frac{\pi}{2}}W(s),
\end{equation}
with a spatially varying range parameter $r(s)$, partitioned into subdomains $\Omega_d$, each with constant range $r_d$. Using finite elements, the SPDE is reformulated in weak form as:

\begin{equation}
\left\langle\psi_j,\left[1 - \nabla\frac{r(\cdot)^2}{8}\nabla\right]\tilde{u}\right\rangle = \left\langle\psi_j, r(\cdot)\sqrt{\frac{\pi}{2}}W(\cdot)\right\rangle.
\end{equation}

The finite element matrices are defined as follows:

\begin{align}
&\mathbf{C}_{i, j}=\left\langle\psi_{i}, \psi_{j}\right\rangle=\int_{\Omega} \psi_{i}(s) \psi_{j}(s) \partial s \notag \\
& \left(\mathbf{G}_{q}\right)_{i, j}=\left\langle 1_{\Omega_{q}} \nabla \psi_{i}, \nabla \psi_{j}\right\rangle=\int_{\Omega_{q}} \psi_{i}(s, s) \partial s \notag \\
& \left(\tilde{\mathbf{C}}_{q}\right)_{i, i}=\left\langle 1_{\Omega_{q}} \psi_{i}, 1\right\rangle=\int_{\Omega_{q}} \psi_{i}(s, s) \partial s
\label{eq:fem}
\end{align} 
with $\mathbf{C}$ computed over the whole domain, while $\mathbf{G}_{q}$ and $\tilde{\mathbf{C}}_{q}$ are defined as a pair of matrices for each subdomain. Then the precision matrix is
\begin{equation}
\mathbf{Q}=\frac{1}{\sigma^{2}} \mathbf{R} \tilde{\mathbf{C}_r}^{-1} \mathbf{R} \text { for } \mathbf{R}_{r}=\mathbf{C}+\frac{1}{8} \sum_{q=1}^{k} r_{q}^{2} \mathbf{G}_{q}, \quad \tilde{\mathbf{C}}_{r}=\frac{\pi}{2} \sum_{q=1}^{k} r_{q}^{2} \tilde{\mathbf{C}}_{q}
\end{equation}

In the case when $r=r_{1}=r_{2}=\ldots=r_{k}$ we have $\mathbf{R}_{r}=\mathbf{C}+\frac{r^{2}}{8} \mathbf{G}$ and $\tilde{\mathbf{C}}_{r}=\frac{\pi r^{2}}{2} \tilde{\mathbf{C}}$ giving

\begin{equation}
\mathbf{Q}=\frac{2}{\pi \sigma^{2}}\left(\frac{1}{r^{2}} \mathbf{C} \tilde{\mathbf{C}}^{-1} \mathbf{C}+\frac{1}{8} \mathbf{C} \tilde{\mathbf{C}}^{-1} \mathbf{G}+\frac{1}{8} \mathbf{G} \tilde{\mathbf{C}}^{-1} \mathbf{C}+\frac{r^{2}}{64} \mathbf{G} \tilde{\mathbf{C}}^{-1} \mathbf{G}\right)
\end{equation}
which coincides with the stationary case in \cite{lindgren_bayesian_2015}, since for $k=1$, $\tilde{\mathbf{C}}=\mathbf{C}$.\\ \\
In practice we can define $p_d$ as the range fraction such that $r_{q}=p_{q} r$, with $p_{1}, \ldots, p_{k}$ with $0 < p \leq 1$ known constants and by default $p_1 = 1$. This gives

\begin{equation}
\tilde{\mathbf{C}}_{r}=\frac{\pi r^{2}}{2} \sum_{q=1}^{k} p_{q}^{2} \tilde{\mathbf{C}}_{q}=\frac{\pi r^{2}}{2} \tilde{\mathbf{C}}_{p_{1}, \ldots, p_{k}} \text { and } \frac{1}{8} \sum_{q=1}^{k} r_{q}^{2} \mathbf{G}_{q}=\frac{r^{2}}{8} \sum_{q=1}^{k} p_{q}^{2} \tilde{\mathbf{G}}_{q}=\frac{r^{2}}{8} \tilde{\mathbf{G}}_{p_{1}, \ldots, p_{k}}
\end{equation}
where $\tilde{\mathbf{C}}_{p_{1}, \ldots, p_{k}}$ and $\tilde{\mathbf{G}}_{p_{1}, \ldots, p_{k}}$ are pre-computed.\\ \\
The main result is summarized in the following proposition.
\begin{proposition}
A non-stationary Matérn-type Gaussian field defined on $\Omega\subset\mathbb{R}^2$ consisting of a system of stationary Matérn SGFs on an exhaustive partition $\Omega_1,\Omega_2,...,\Omega_k\subset \mathbb{R}^2$ with spatially varying range parameters $r,p_2r,...,p_kr>0$ and marginal standard deviation $\sigma_u$ is defined by the following log density function
\begin{equation}
    \log q(u|\sigma_u,r,p_1=1,p_2,...,p_k)\propto u^\top \pmb Q_u u
\end{equation}
where
$\pmb{Q}_u=\frac{1}{\sigma^{2}} \mathbf{R} \tilde{\mathbf{C}_r}^{-1} \mathbf{R} \text { for } \mathbf{R}_r=\mathbf{C}+\frac{1}{8} \sum_{q=1}^{k} p_{q}^{2}r^2 \mathbf{G}_{q}, \quad \tilde{\mathbf{C}}_{r}=\frac{\pi}{2} \sum_{q=1}^{k} p_{q}^{2}r^2 \tilde{\mathbf{C}}_{q}$, and $\pmb C, \pmb G_q, \tilde{\pmb C}_q, q = 1,2,...,k$ is calculated according to \eqref{eq:fem}. This model is termed the Transparent Barrier model (TBM). \\
Thus, the prior for $u(s)$ is given as
\begin{equation*}
u(s)|\sigma_u,r,p_1,p_2,...,p_k \sim N(0, \pmb Q_u^{-1}).
\end{equation*}
Note that if $k=1$, the TBM reduces to the stationary Matérn SGF and for $k=2$ with $p_2$ very small, the TBM reduces to the barrier model of \cite{bakka_non-stationary_2019} with one impermeable barrier.
\label{prop}
\end{proposition}

\begin{figure}[h]
\centering
\includegraphics[width=0.9\textwidth]{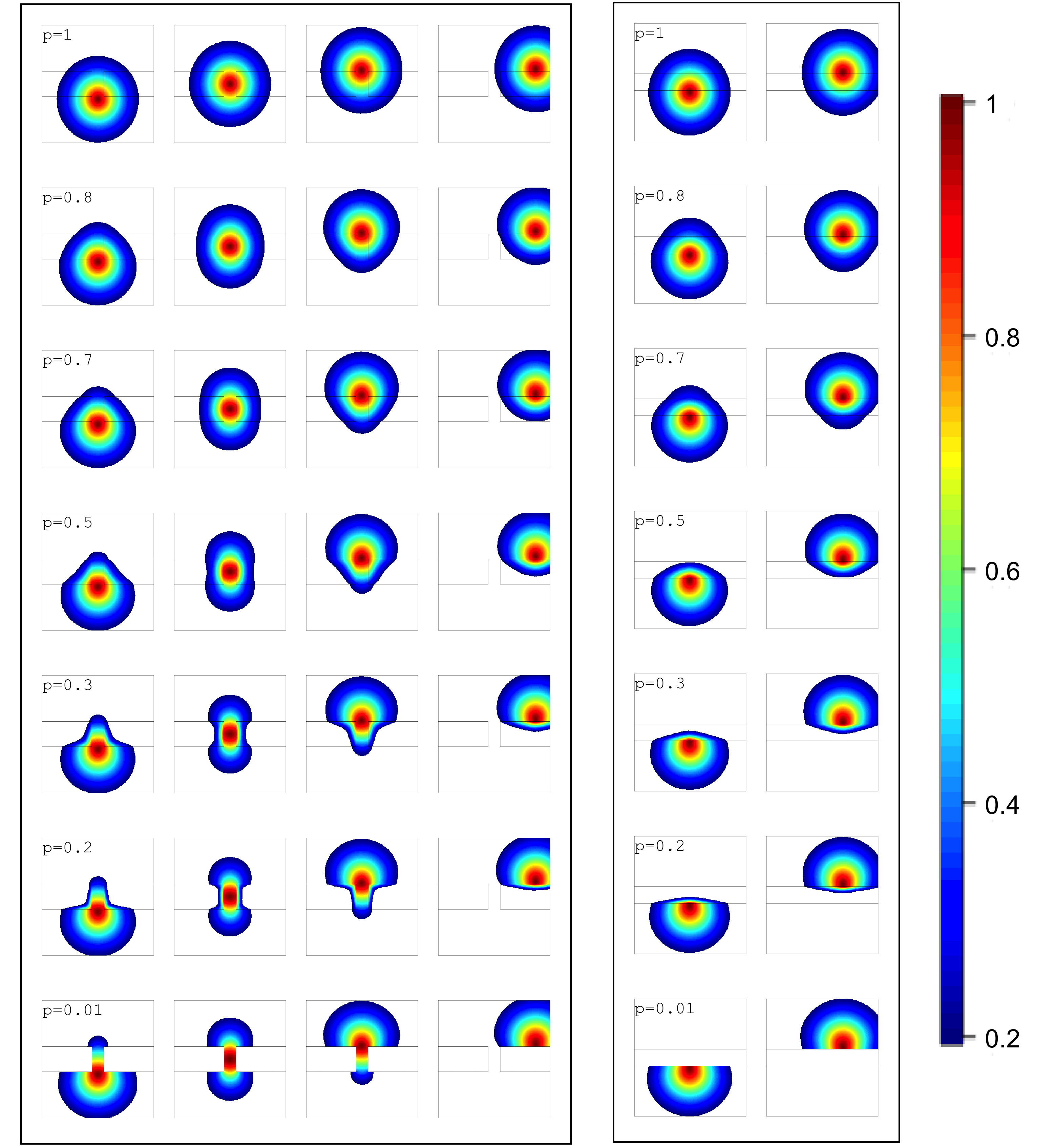}
\caption{Correlation plots illustrating spatial dependence for two barrier configurations. The first configuration (left side) represents a study area divided by a barrier with a canal in the middle, connecting the upper and lower sections of the normal area. The second configuration (right side) depicts a study area completely divided by a thin barrier. Rows show prior correlation obtained with range fraction: 1 (row 1), 0.8 (row 2), 0.7 (row 3), 0.5 (row 4), 0.3 (row 5), 0.2 (row 6), and 0.01 (row 7). Columns show distinct reference points 
for which spatial correlation is plotted, these points are all in the normal area.}\label{fig2}
\end{figure}

Transparent barrier permeability is determined by the barrier range fraction, defined as a fraction of the spatial range parameter in the normal area. Higher range fraction values increase barrier permeability, thereby strengthening spatial connections across barriers. In Figure \ref{fig2} the correlation between a point and all other points is shown for various scenario's of the TBM. We note how the barriers transition from impermeable to completely permeable as the range fraction increases. Specifically, the first row of Figure \ref{fig2} corresponds to a fully impermeable scenario, which can be modeled using the existing Barrier model, whereas the last row corresponds to the stationary model scenario where the barriers are ignored. \\ \\
When encountering a barrier, the distribution of possible locations for a moving point is influenced by the barrier's permeability. Under the Barrier model, the probability of finding the point on the opposite side of the barrier is nearly non-existent (Figure \ref{fig2}, top right) unless there is a path around the barrier, such as a canal or gap, where some correlation may still occur (Figure \ref{fig2}, top left). In contrast, under a stationary Gaussian random field, the spatial correlation structure remains unaffected by any barriers, exhibiting consistent correlations regardless of the barrier's presence or the point's location within the study area (bottom row of Figure \ref{fig2}). The TBM aims to distort the spatial correlation that the moving point would exhibit under a stationary scenario, but without reaching the extreme behavior of the fully impermeable scenario. This distortion, controlled by adjusting the fraction of the range parameter used in the barrier area, depends on the specific characteristics and nature of barriers encountered in real-world applications.

\subsection{Transparency Definition}

A central aspect of the Transparent Barrier model is its capability to systematically adjust barrier permeability by modifying the local spatial range parameter in each region. Practically, this is implemented by expressing the range within a barrier as a fraction of the range in the normal (non-barrier) area, where lower fractions represent less permeable barriers. While the ecological motivation is clear, translating this into a parametrization for the range fraction requires some consideration. \\ \\

Let $d_n$ be the distance in number of units at which correlation is $c_0$ in the normal area, and $d_b$ be the distance in number of units at which correlation equal $c_0$ inside (any) barrier. We then define transparency as:

\begin{equation}
t_{c_0} = \frac{d_b}{d_n}|_{c_0}
\end{equation}

where $t_{c_0}$ is the \textbf{transparency} evaluated at the chosen reference correlation $c_0$, thus $t_{c_0}$ is a fixed fraction of the distance at which correlation is $c_0$.

For the specific case where $c_0 \sim 0.13$, the comparison between the distance at which the correlation is $0.13$ inside and outside is nothing more than what we already know to be the range fraction, then $t_{c_0}=p_b = r_b/r_n$ ($t_{c_0=0.13} = p_b$ in Table \ref{tab1}). 
  
To find $t_{c_0}$ we: (1) fit a spline $f(d)$ for the curve distance over correlation obtained with the Transparent Barrier model assuming the range to be the prior range for the normal area in the entire study region. (2) fit a scaled version of the spline $f_{scaled}(d)=f(d/s)$ with scaling factor $s$. And (3) find $s$ so that:

\begin{equation}
f_{scaled}(d_b) = c_0 \text { and } t_{c_0} = \frac{d_b}{d_n}|_{c_0}
\end{equation}
Then $p_b = r_b/r_n$, with $r_n$ the prior range of the normal area and  $r_b$ the distance at which correlation is 0.13 on the scaled function. The scaling factor $s$ represents the relative rate at which spatial correlation decays in a hypothetical scenario where the range across the entire domain is set to $r_b$. \\ \\
While permeability is explicitly defined in the model by the range fraction of a barrier $p_b$, transparency $t_{c_0}$ is a way of considering previous knowledge we have about the study area and use it to find the range fraction.\\ \\
Table \ref{tab1} shows some examples of $t_{c_0}$ with $c_0$ equal to 0.13, 0.5, and 0.8. See Appendix \ref{secA1} for more details on splines, and code used to get Table \ref{tab1}.
\begin{table}[h]
\caption{Examples of transparencies $t_{c_0}$ at various correlation values $c_0$, the corresponding scaling factors $s$ used, and the resulting range fractions $p_b$ implemented in the Transparent Barrier model.}\label{tab1}
\begin{tabular*}{\textwidth}{@{\extracolsep\fill}lcccccc}
\toprule%
& \multicolumn{2}{@{}c@{}}{$c_0 = 0.13$} & \multicolumn{2}{@{}c@{}}{$c_0 = 0.5$} & \multicolumn{2}{@{}c@{}}{$c_0 =0.8$}  \\\cmidrule{2-3}\cmidrule{4-5}\cmidrule{6-7}%
$t_{c_0}$ & $s$ & $p_b$ & $s$ & $p_b$ & $s$ & $p_b$ \\
\midrule
0.1  & 7.04  & 0.1 & 1.95  & 0.730 & 1.24 & 0.914 \\
0.2  & 6.03  & 0.2 & 1.86  & 0.748 & 1.22 & 0.921 \\
0.3  & 4.99  & 0.3 & 1.76  & 0.771 & 1.20 & 0.928 \\
0.4  & 4.06  & 0.4 & 1.65  & 0.796 & 1.18 & 0.936 \\
0.5  & 3.28  & 0.5 & 1.53  & 0.829 & 1.15 & 0.944 \\
0.6  & 2.61  & 0.6 & 1.42  & 0.861 & 1.12 & 0.954 \\
0.7  & 2.07  & 0.7 & 1.30  & 0.895 & 1.09 & 0.965 \\
0.8  & 1.64  & 0.8 & 1.19  & 0.930 & 1.06 & 0.976 \\
0.9  & 1.28  & 0.9 & 1.09  & 0.965 & 1.03 & 0.988 \\
\hline
\end{tabular*}

\end{table}


The results for $c_0 = 0.5$ offer the following empirical interpretation: $1-t_{c_0} = \frac{d_n-d_b}{d_n} |_{c_0}$ which represents the proportional reduction in distance, measured relative to the distance in the normal area $d_n$ at the correlation level $c_0$. When the relationship between distance and correlation has an approximate one to one correspondence, $c_0 \sim 0.5$, the following can be interpreted: to achieve a distance reduction of $1-t_{c_0}$, the smoothing spline must decay at a rate approximately $1+(1-t_{c_0})$ times faster.\\ \\
At $c_0 = 0.5$, the relationship $1-t_{c_0} \sim s-1$ (equivalently $2-t_{c_0} \sim s$) provides a practical approximation. For instance, specifying $t_{0.5} = 0.2$ (Table~\ref{tab1}) indicates a desired correlation decay over a distance that is 80\% of that in the normal area, \textit{i.e.}, $0.8 \times d_n$. To enforce this decay rate, the global decay rate of the correlation function must increase by approximately a factor of $ \sim 1 +0.8$. This implies that correlation decays roughly 80\% faster within the barrier region relative to the normal area. Conversely, if it is intended for the correlation to decay 80\% faster in the barrier, setting $t_{0.5} = 0.2$ provides a direct path to determining the appropriate barrier range parameter $r_b$.\\ \\
In the absence of detailed information on how distance decays within the barrier region relative to the normal area at a fixed correlation level $c_0$, it may be reasonable to base the analysis on $t_{0.5}$ as a default reference point.\\ \\
Alternative interpretations of transparency are possible depending on the choice of $c_0$. For instance, if there is some understanding of movement changes at the edge of a barrier, but little knowledge of movement farther from the edge, selecting a higher $c_0$ would be appropriate. Higher correlation values are associated with shorter distances, and therefore reflect spatial relationships at finer scales near the boundary between normal and barrier areas. Conversely, if the distance at which spatial correlation effectively disappears is known, a lower value of $c_0$ may be chosen, as it corresponds to broader-scale correlation patterns that extend further into the barrier.


\section{Simulation study}\label{sec:sim}
In this section we show that the TBM can capture the complex spatial field generated from areas with spatially varying ranges due to permeable barriers in the study domain. We also show that the stationary and barrier models are the limiting cases of the TBM as presented in Proposition \ref{prop}.

\subsection{Discretization of the study area}

In non-stationary spatial contexts involving physical barriers, the mesh explicitly represents islands and other barriers as internal boundaries, ensuring accurate modeling of varying barrier permeability. Proper boundary specification is critical, as it directly influences the representation of spatial dependencies and model accuracy. Mesh construction, including the adequate representation of boundaries, remains an essential user-assessed step; however, this assessment is similarly required when applying stationary spatial Gaussian fields \cite{krainski_advanced_2018}.

\begin{figure}[h]
\centering
\includegraphics[width=1\textwidth]{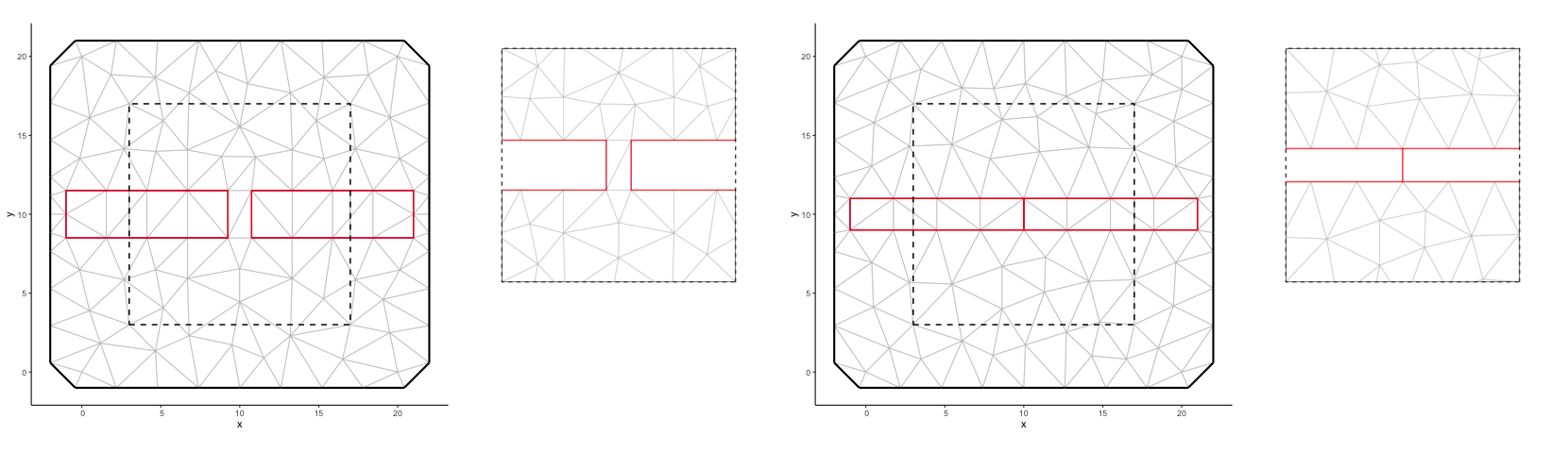}
\caption{Coarse mesh example with barriers in red. The first configuration (left side) represents a study area divided by a barrier with a canal in the middle. The second configuration (right side) depicts a study area divided by a barrier without a canal. Large plots show the extended mesh, and small plots the area of interest only.}\label{fig1}
\end{figure}

The mesh examples constructed placed the outer boundary far from the area of interest to avoid boundary effects \cite{bakka_non-stationary_2019}. Figure  \ref{fig1} shows a coarse version of the mesh used in this simulation study. The study area is enclosed by a dashed line and the barriers areas by a red line. Two barrier configurations are used to illustrate the mesh and following simulations. The first configuration represents a study area divided by a barrier that has a canal in the middle connecting the top and bottom of the normal area. The second configuration represents a study area completely divided by a thin barrier.

\subsection{Model}

To evaluate model performance, spatial fields were simulated using the TBM and then fitted using three distinct approaches: the \textbf{stationary model}, assuming no barriers; the \textbf{Barrier model}, assuming fully impermeable barriers; and the proposed \textbf{Transparent Barrier model}, allowing varying permeability. This evaluation framework assesses each model's effectiveness in recovering the underlying spatial field under different permeability scenarios.

\begin{figure}[h]
\centering
\includegraphics[width=0.9\textwidth]{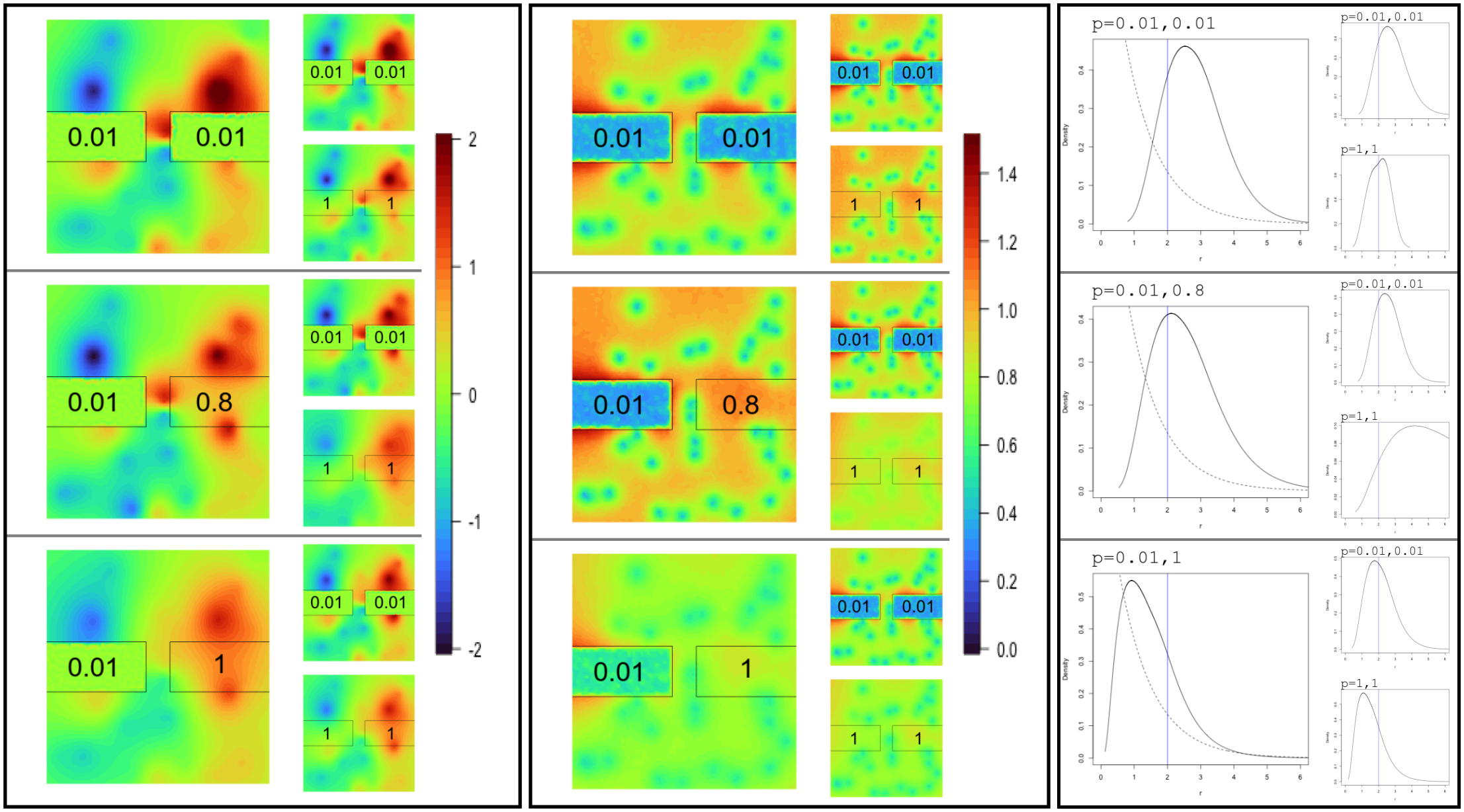}
\caption{
Posterior results for the configuration with two connected barriers and no canal. Each row with bigger plots represents a different range fraction (0.01, 0.8, 1) used to simulate the permeability of the barrier on the right, with the left barrier remaining fully impermeable. Sets of plots, from left to right: (set to the left) the posterior mean of the spatial field, (set in the middle) the posterior standard deviation, and (set to the right) the posterior distribution of the spatial range in the normal area. On each setting, the Transparent Barrier model is displayed in large plots on the left, followed by smaller plots for the Barrier model (top) and stationary model (bottom).}\label{fig3}
\end{figure}

\begin{figure}[h]
\centering
\includegraphics[width=1\textwidth]{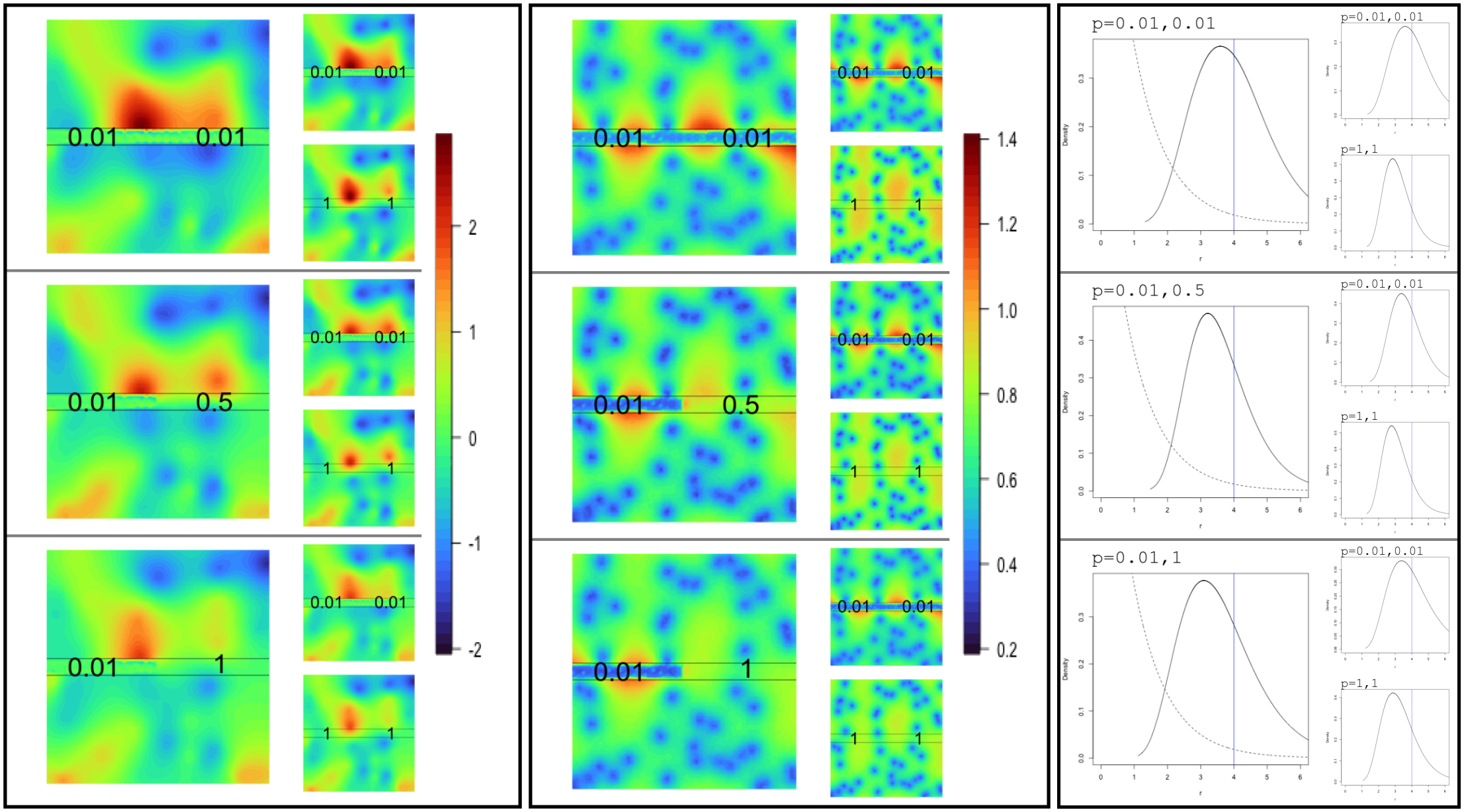}
\caption{Posterior results for the configuration with two connected barriers and no canal. Each row with bigger plots represents a different range fraction (0.01, 0.5, 1) used to simulate the permeability of the barrier on the right, with the left barrier remaining fully impermeable. Sets of plots, from left to right: (set to the left) the posterior mean of the spatial field, (set in the middle) the posterior standard deviation, and (set to the right) the posterior distribution of the spatial range in the normal area. On each setting, the Transparent Barrier model is displayed in large plots on the left, followed by smaller plots for the Barrier model (top) and stationary model (bottom).}\label{fig4}
\end{figure}

Two geometric configurations were used in the simulations: one with the study area divided by two barriers with a connecting canal (Configuration 1), and another where the area is divided by two connected barriers without a canal (Configuration 2). These configurations are illustrated in Figure \ref{fig3} and Figure \ref{fig4}, respectively.

Spatial ranges of 2 (Configuration 1) and 4 (Configuration 2) were set for the normal area, and and a wide range of barrier permeability scenarios, represented by range fractions from 0.01 (nearly impermeable) to 1 (fully permeable), were evaluated. Specific results for range fractions of 0.01, 0.1, 0.3, 0.4, 0.5, 0.7, 0.8, and 1 are detailed in the supplementary material.

Thus, the data, $y(s_i), i=1,2,...,n,$ were simulated according to either Configuration 1 or Configuration 2
as follows:
\begin{eqnarray*}
u|\sigma_u, r,p_1=1,p_2&\sim& N(0, \pmb Q)\text{ from Proposition } \ref{prop}\\
    Y|u,\sigma_y&\sim& N(u,\sigma_y^2).
\end{eqnarray*}

\subsection{Results}
The posterior spatial field shows clear differences across models. At low permeability (e.g., range fraction = 0.01), the Transparent Barrier model matches the Barrier model, with a sharp discontinuity across the barrier. As permeability increases, the Transparent Barrier model captures more spatial continuity across the barrier, unlike the Barrier model, which maintains discontinuity regardless of the true range fraction used for the simulation, and the stationary model, which overly smooths across barriers. When the right barrier’s range fraction is close to 1, the Transparent Barrier model closely resembles the Stationary model but remains distinct due to the impermeable left barrier.

Posterior standard deviations follow a similar trend. As permeability increases, the Transparent Barrier model distributes uncertainty across the barrier more realistically than the other two models, which either overstate the standard deviation (Barrier) or understate it (stationary), especially near barrier edges. Higher standard deviation near the barrier when permeability is low can be understood by analogy with a moving individual who becomes ``trapped'' upon reaching a boundary, limiting their potential movement and resulting in longer residence times near the barrier relative to more open regions. Moreover, individuals are far less likely to cross the barrier than to move along the same region (either the upper or lower side of the study area). When the range fraction is close to 0.01, crossing is effectively impossible. Uncertainty is then explained because the concentration of individuals at the edge of the barrier is either very high or very low.

Regarding the posterior distribution of the spatial range, the stationary model tends to underestimate the range in the normal area because it does not distinguish between regions of differing permeability. By averaging over both the barrier and normal areas, despite the lower correlation induced by the barrier, the model produces a posterior estimate biased toward a lower range. Conversely, the Barrier model overestimates the range, likely compensating for its assumption of impermeability. 


These findings demonstrate the Transparent Barrier model’s effectiveness in bridging the gap between impermeable and stationary assumptions, providing more realistic spatial inference in domains where barriers of varying permeability are present.

\section{Application to Dugong species distribution in the Red Sea}\label{sec:app}

The Arabian Red Sea, a marine biodiversity hotspot and home to vulnerable dugong populations with limited data \cite{preen_distribution_2004, nasr_status_2019, khamis_identifying_2022}, provides an ideal context for demonstrating the Transparent Barrier model. \\ \\
We modeled dugong distribution using incidental sighting data collected opportunistically through tourism activities, research expeditions, and citizen science initiatives along the northern coast of the Saudi Arabian Red Sea, an area characterized by a complex spatial structure. Existing geographic data often group islands imprecisely, making detailed spatial modeling challenging. To address this, we used bathymetry data not as a direct environmental covariate, but to more accurately represent the location and depth of islands, despite its coarse 100-meter resolution. We adopted the log-Gaussian Cox process model (LGCP), which is well-suited for modeling a point pattern \cite{simpson_going_2016, illian_toolbox_2012}.\\ \\
The LGCP is a well-known stochastic process often used to model occurrence-only data observed as a point pattern, for instance the locations of earthquakes, certain specie of animal, plants, disease cases and so on. The LGCP is a non-homogenous Poisson process \cite{moller_statistical_2003, moraga_spatial_2023, simpson_going_2016} where the log intensity function is modeled as a latent Gaussian process. Thus we observe locations $s_i\in \mathbb{R}^2, i = 1,2,...,n$ over a study domain $\Omega$. The LGCP is defined based on a SGF $u(s)$ as
\begin{equation}
    Y|\lambda \sim \text{Poisson}(\lambda)\quad
    \log\lambda(s) = \beta^\top \pmb X + u(s)\quad
    u(s) \sim N\left(0, \pmb\Sigma_u\right),
\end{equation}
thus for a certain area $A \subset \Omega\subset\mathbb{R}^2$, the expected number of occurrences follows a Poisson distribution with expected value $\lambda_a = \int_A\lambda(s)ds$, such that $Y_A|\lambda \sim \text{Poisson}(\int_A\lambda(s)ds)$. \\
If we use the TBM as defined in Proposition \ref{prop} for $u(s)$, then we can model a point pattern as an LGCP in the presence of permeable and impermeable barriers. In our case we have no covariates for fixed effects and only include an overall intercept so that the TBM captures the deviations from the overall intensity.

 The incidental sightings of dugongs and the bathymetry data were provided by the Red Sea Zone Department of Environmental Protection and Regeneration (RSZ-DEPR). Dugong locations have been adjusted slightly to maintain confidentiality without compromising the conclusions of our study; however, these adjusted locations should not be used by others as an accurate representation of dugong distribution.

\subsection{Mesh construction}

\begin{figure}[h]
\centering
\includegraphics[width=1\textwidth]{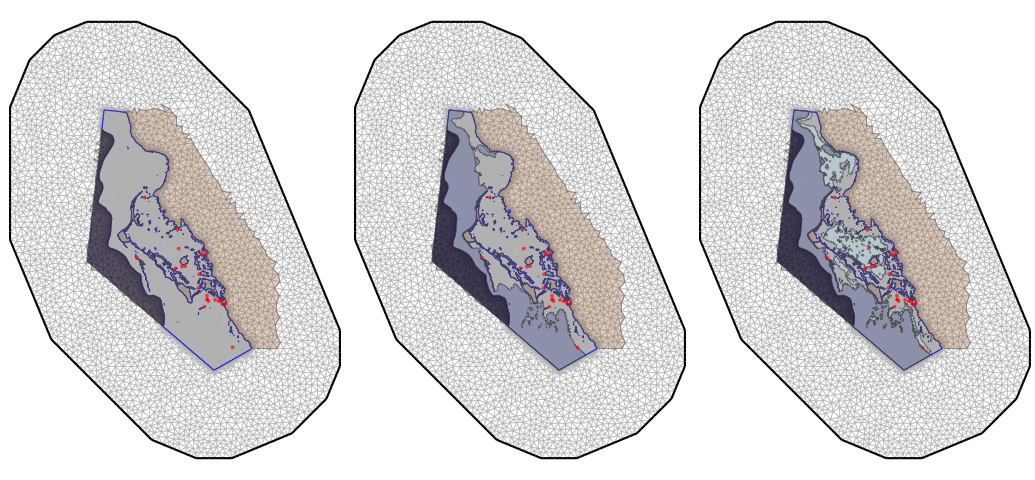}
\caption{Mesh used for the application example, with triangle centers classified into normal area or barriers. Left: Barrier 1, including land (light brown) and sea deeper than –500 meters (dark blue). Middle: Barrier 2 (gray), corresponding to depths between –500 and –50 meters, added to the previous configuration. Right: Barrier 3 (light blue), representing depths between –50 and –30 meters, added on top of barriers 1 and 2. Red dots indicate Dugong observations.}\label{fig5}
\end{figure}

We added boundaries for impermeable and permeable barriers in three steps. We defined the first boundary primarily prior to the mesh construction, and defined the second and third boundaries over the mesh.\\ \\
First, we delineated an initial boundary defining the area of interest and land within based on the extent of the bathymetry map and areas with bathymetry greater than or equal to 0. To this set of boundaries, we added sea deeper than 500 meters and considered all coastlines, islands, and sea deeper than 500 meters as \textit{holes} for mesh construction. 
Then, we partitioned the domain using a Delaunay triangulation, with triangle size adapted to spatial complexity: finer near observation-dense areas and around barriers, coarser elsewhere. We included an external buffer outside the delineated study area in the mesh construction to minimize edge effects. Continental land on the right side of the study area was part of this buffer zone. Continental land, coastlines, islands, and sea deeper than 500 meters were considered impermeable physical barriers, and we refer to these as barrier 1 (Figure \ref{fig5}, left plot).\\ \\
For the second and third steps, we added permeable barriers 2 and 3, respectively. Barrier 2 was defined over the area with depths between 500 and 50 meters (Figure \ref{fig5}, middle plot), and barrier 3 over the area with depths between 50 and 30 meters (Figure \ref{fig5}, right plot). These two barriers were treated differently in the model, as we assigned distinct range fractions to each in order to reflect assumed differences in permeability. Although dugongs can theoretically be found at both of these depths, their diving behavior and the presence of predators in deeper waters suggest they are far less likely to use these areas than shallower ones. The deeper barrier (barrier 2) is located in the open sea, where crossing provides little benefit to a dugong. In contrast, the shallower barrier (barrier 3), although still deep, lies within a semi-enclosed area surrounded by islands and shallower coastlines, suggesting it may occasionally be crossed to move between foraging areas.

\subsection{Results}

\begin{figure}[h]
\centering
\includegraphics[width=1\textwidth]{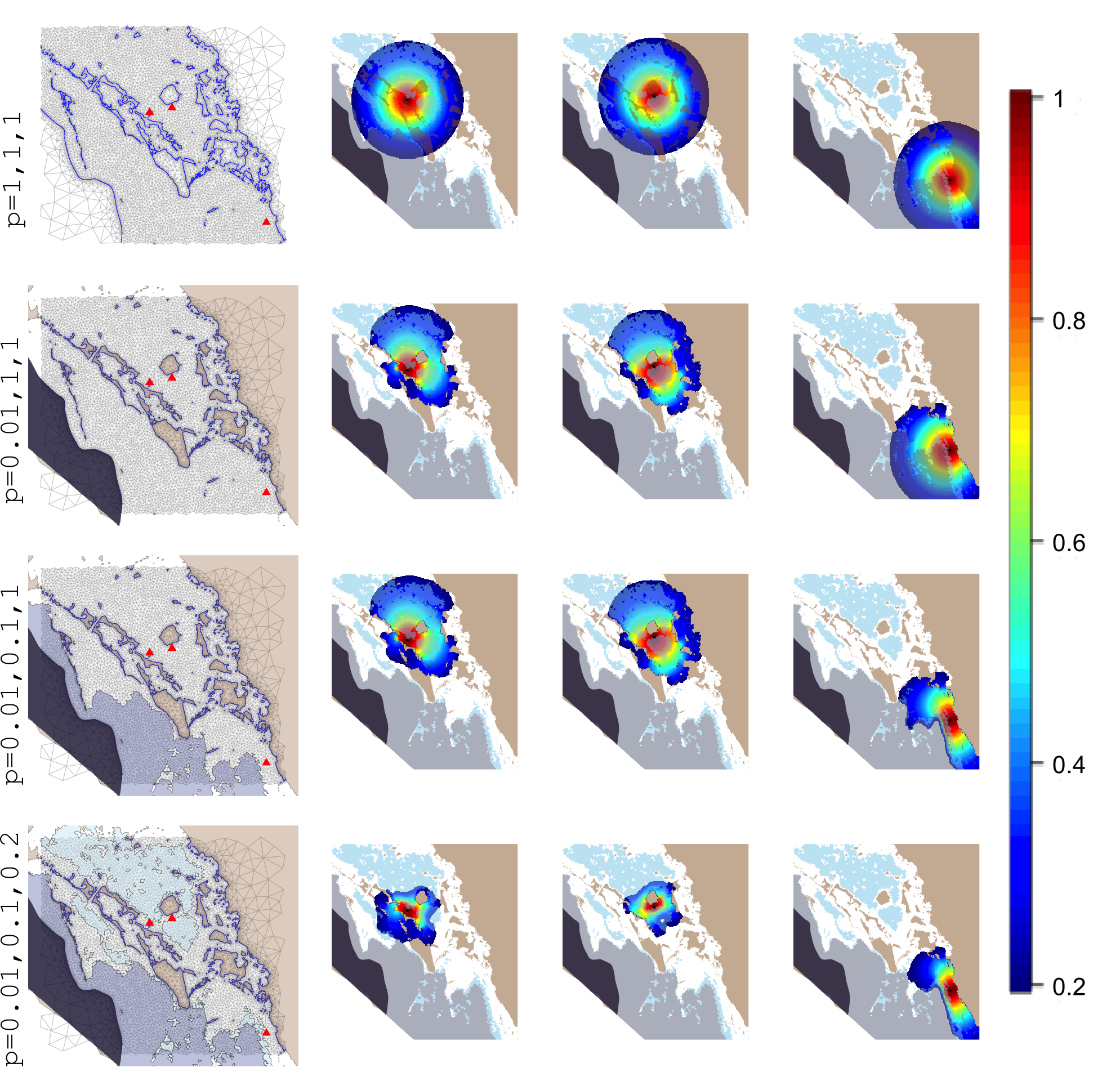}
\caption{Prior correlation surfaces under varying barrier permeability assumptions. Each row corresponds to a different configuration of range fractions for barriers: (row 1) all barriers have range fraction 1; (row 2) barrier 1 has range fraction 0.01, others 1; (row 3) barrier 1 = 0.01, barrier 2 = 0.1, barrier 3 = 1; (row 4) barrier 1 = 0.01, barrier 2 = 0.1, barrier 3 = 0.2. The first column shows the mesh with reference points in red used to generate correlation plots shown in columns 2–4. Colored regions indicate barrier types: light brown (land), dark blue (sea $<$ -500 m), gray (-500 to -50 m), light blue (-50 to -30 m). This figure is a close up of the entire study area.}\label{fig6}
\end{figure}

Before data fitting, we assessed barrier design impacts on spatial structures using prior correlation. We evaluated four models: no barriers, barrier 1 only, barriers 1 and 2, and all three barriers. In the first model, all range fractions were set to 1, meaning the model effectively became a stationary spatial field. This  case  exhibited no discontinuities in correlation, allowing spatial dependence to propagate over land (Figure \ref{fig6}, row 1). While unrealistic for marine species like dugongs, we used it to answer whether using a non-stationary model is justifiable.\\ \\
The second model considered barrier 1 impermeable by assigning a range fraction of 0.01 As a result, spatial dependence is appropriately constrained to follow plausible pathways in water (Figure \ref{fig6}, row 2). Although, this setup still assumes permeability in open and semi-enclosed deep areas, which may lead to unrealistic movement in regions where dugongs are unlikely to travel, it serves as a benchmark to evaluate the effects of adding transparent barriers to the model.\\ \\
The third model had a more realistic representation. This model considered barrier 1 to be impermeable and assigned a range fraction of 0.1 to barrier 2 (Figure \ref{fig6}, row 3). This resulted in visibly attenuated correlation across open water, especially noticeable near the southern portion of the domain, where barrier 2 and normal areas were adjacent. The impact was less evident where normal area is enclosed by islands as correlation is cut off by barrier 1 and not by the effect of barrier 2. \\ \\
Finally, the most comprehensive model incorporated barrier 3 in addition to barrier 1 and barrier 2, by assigning a range fraction of 0.2 to the barrier 3 area (Figure \ref{fig6}, row 4). This structure reflects ecological expectations based on dugong diving behavior: in the area enclosed by islands some movement is possible in deep water, but with a clear preference for shallow coastal areas. This behavior is consistent with the motivation for introducing transparent barriers, offering a flexible compromise between unrestricted and entirely blocked movement.

\begin{figure}[h]
\centering
\includegraphics[width=0.7\textwidth]{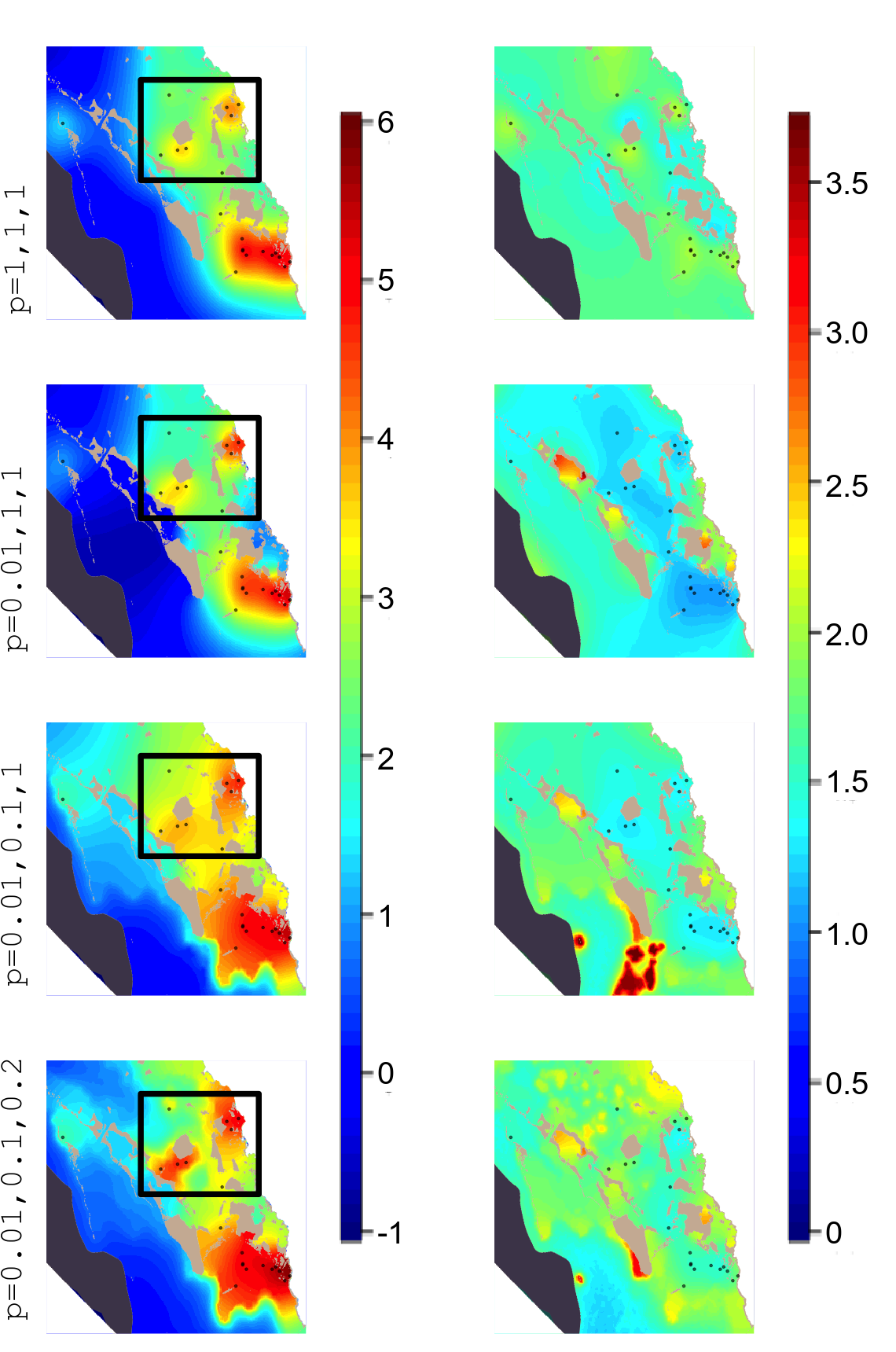}
\caption{Posterior mean (left column) and posterior standard deviation (right column) of the spatial effect under different barrier configurations. Each row corresponds to a barrier setting: (row 1) all barriers have range fraction 1; (row 2) barrier 1 has range fraction 0.01, others 1; (row 3) barrier 1 = 0.01, barrier 2 = 0.1, barrier 3 = 1; (row 4) barrier 1 = 0.01, barrier 2 = 0.1, barrier 3 = 0.2. Only impermeable barriers are colored with light brown indicating land, and dark blue sea deeper than 500 meters. Black dots represent dugong observations. This figure is a close up of the entire study area. The are enclosed in the square is presented in Figure \ref{fig:zoom}.}\label{fig7}
\end{figure}

\begin{figure}[h]
\includegraphics[width=1\textwidth]{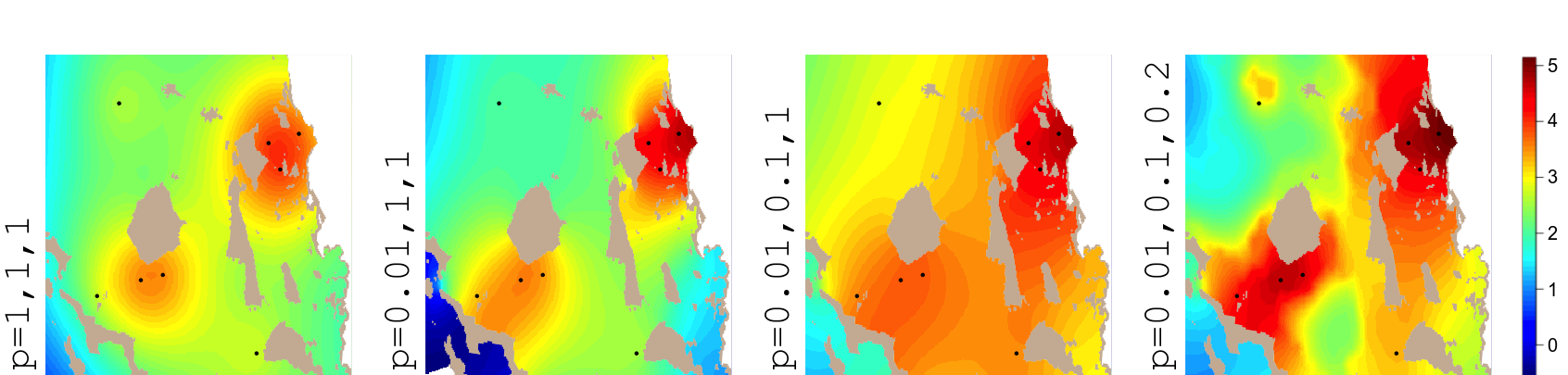}
\caption{Posterior mean of the spatial effect under different barrier configurations for the area indicated by the square in Figure \ref{fig7}. From left to right: Stationary model; 
Barrier 1 has range fraction 0.01, others 1; Barrier 1 = 0.01, Barrier 2 = 0.1, Barrier 3 = 1; Barrier 1 = 0.01, Barrier 2 = 0.1, Barrier 3 = 0.2. Only impermeable barriers are colored with light brown indicating land, and dark blue sea deeper than 500 meters. Black dots represent dugong observations.}
\label{fig:zoom}
\end{figure}

Under the same four models, we examined the posterior distribution of the mean and standard deviation of the spatial field (Figure \ref{fig7}, column 1 and column 2 respectively). In the absence of any barriers (Figure \ref{fig7}, row 1), the posterior mean exhibited a smooth spatial gradient: high where observations are located, and gradually decreasing away from these areas. However, it remained unrealistically positive in areas with no observations where access is difficult due to surrounding islands. Introducing barrier 1 (Figure \ref{fig7}, row 2) caused visible changes in the posterior mean along the coastline. There were sharp transitions in the mean of two adjacent areas if there was land in between, and we observed localized increases in mean values near barrier edges, where the spatial field concentrates due to restricted movement pathways. Adding barrier 2, representing deep open sea (Figure \ref{fig7}, row 3), resulted in high mean values concentrated in the area enclosed by islands. Incorporating barrier 3 (Figure \ref{fig7}, row 4) segments the space in this semi-enclosed area, and high mean values became more localized. We observed that connections across the semi-enclosed area were strengthened or weakened depending on water depth. As dugong movement becomes more constrained in deep water, coastal areas with no observations have higher mean values compared to the second model with barrier 1 only. This redistribution aligns with realistic movement expectations given dugong diving behavior and avoidance of deep water.\\ \\
As we observed in the simulation study, barriers increased local uncertainty by limiting spread. Posterior standard deviations for the field showed that in narrow inlets or barrier-enclosed zones, uncertainty rises due to the difficulty for dugongs to access or exit these zones; thus, the presence of dugongs is either high or very low. We also observed that barrier presence produces a patchier uncertainty landscape (Figure \ref{fig7}, column 2).\\ \\
In Figure \ref{fig:zoom} we focus on the most northern part of the study area. This area is interesting because we have only three sightings here. When we do not account for the barriers (most left in Figure \ref{fig:zoom}), we predict a higher intensity only around the observed sightings. As we account for Barrier 1, Barrier 2 and Barrier 3 (from left to right in Figure \ref{fig:zoom}) the predicted intensity increases in the northern coastal areas due to our model borrowing strength from other similar areas with more sightings and the spatial correlation through the canals. This area would not be identified as a focus area for conservation interventions under the stationary model, but under the TBM we see that this area is important for Dugong species distribution.\\ \\
These results affirm the Transparent Barrier model’s ability to reflect ecological constraints and discontinuities in spatial dependence, generating results that align with known habitat preferences and movement restrictions in shallow-water marine species like the dugong.

\section{Discussion}\label{sec:disc}

Realistic representation of spatial processes is crucial for ecological modeling, especially when physical barriers significantly influence species distribution. Traditionally, species distribution models (SDMs) have employed methods such as generalized additive models, random forests, or maximum entropy models, largely dependent on comprehensive environmental data availability \cite{reiss_species_2011, roberts_habitat-based_2016, wang_mapping_2020}.
These traditional approaches assume spatially uncorrelated data, which makes them unsuitable for marine megafauna populations with low individual counts, such as dugongs.\\ \\
Modern approaches accounting for spatial dependence frequently assume spatial stationarity and isotropy, which are often unrealistic in the presence of physical barriers such as islands or coastlines. The Barrier model introduced by \cite{bakka_non-stationary_2019} addresses this issue by explicitly modeling impermeable barriers, but is limited by its assumption that all barriers are completely impermeable, overlooking real-world scenarios with varying permeability. To overcome these limitations, we introduced the Transparent Barrier model, a novel approach which enables the representation of barriers of varying permeability. Unlike previous barrier models, this method can realistically represent scenarios where species movements are possible but limited by permeable barriers. Consequently, it improves the flexibility and applicability of spatial modeling and can be used to provide more accurate ecological insights.\\ \\
The results of our simulation study underscore the value of the Transparent Barrier model in capturing complex spatial structures shaped by natural obstacles. It consistently outperformed both the Barrier model and the stationary model in recovering underlying spatial structures, particularly around the edge of the barrier, across various geometrical configurations and permeability scenarios. Moreover, this modeling approach maintains computational efficiency, ensuring rapid implementation using the integrated nested Laplace approximation (INLA). Hence, the Transparent Barrier model is both theoretically robust and practically advantageous for real-world ecological modeling.\\ \\
Applying our model to Dugong species distribution along the northern coast of the Saudi Arabian Red Sea yielded insights into the influence of physical barriers on species distribution. Our results align with existing literature, where dugongs exhibit complex movement patterns influenced by environmental factors, such as bathymetric depth, proximity to shore, predator presence, and human activities \cite{wirsing_behavioural_2012, seal_spatial_2024, shawky_dugong_2024}, and also identifies areas that are crucial for conservation efforts. The Transparent Barrier model effectively captured these nuances, depicting realistic patterns of dugong habitat use, particularly in the restricted use of deeper waters as barriers to movement, consistent with their known shallow diving behavior \cite{chilvers_diving_2004, sheppard_dugong_2010}.\\ \\
The applicability of the Transparent Barrier model can be extended beyond dugong conservation. Marine environments, characterized by complex coastlines, islands, and bathymetric features, commonly experience substantial spatial variability in species distributions due to physical barriers. Fisheries research extensively emphasizes the importance of such barriers, highlighting how physical constraints impact fish population dynamics, dispersal, and conservation management \cite{letcher_population_2007, nislow_variation_2011, pepino_fish_2012}. Our model is particularly relevant for marine megafauna conservation, as large marine species typically exhibit extensive and varied spatial movement behaviors influenced significantly by physical and ecological barriers \cite{mcclellan_understanding_2014-1, pimiento_functional_2020}.\\ \\
Some practical and methodological challenges remain. First, our model's accuracy inherently depends on the quality and resolution of available spatial data. For instance, the coarse resolution of the bathymetric data utilized may affect the precise representation of barrier permeability and consequently the accuracy of predictions. Additionally, our incidental sightings dataset, derived from citizen science and opportunistic observations, may introduce biases due to uneven sampling effort. Moreover, the model currently assumes spatially uniform permeability within designated barrier regions, potentially oversimplifying real-world spatial dynamics where permeability might vary continuously. Consequently, future developments of the Transparent Barrier model should explore continuous permeability gradients within barrier regions to reflect even more realistic spatial interactions.\\ \\
Future research should aim at integrating higher-resolution spatial datasets, advanced observational technologies such as unmanned aerial vehicles (UAVs), and telemetry data for a more precise characterization of species distributions. Additionally, extending the Transparent Barrier model to spatiotemporal settings would enable capturing temporal variations in barrier permeability and species movements, further enhancing ecological insights and conservation effectiveness.\\ \\
The transparent barrier model is implemented in the R package \textit{INLAspacetime}.

\section{Appendix A. Transparency values}\label{secA1}

\subsection{Splines}

To estimate a smooth function \( f(x) \) given data \( (x_i, y_i) \), define the penalized criterion:
\[
J(f) = \sum_{i=1}^n [y_i - f(x_i)]^2 + \lambda \int_{x_1}^{x_n} [f''(x)]^2 \, dx
\]

where the first term measures closeness to the data and the second penalizes curvature, with smoothing parameter $\lambda \geq 0$. When $\lambda = 0$ the solution interpolates all points; as $\lambda \to \infty$ it nearly becomes linear. The smoothing parameter \( \lambda \geq 0 \) controls the trade-off:
\[
\lambda = 0 \Rightarrow \text{interpolation}, \quad \lambda \to \infty \Rightarrow \text{linear regression}
\]

The minimizer \( f(x) \) is a natural cubic spline with knots at the \( x_i \), represented piecewise as:
\[
f(x) =
\begin{cases}
S_1(x), & x_1 \leq x < x_2 \\
S_2(x), & x_2 \leq x < x_3 \\
\quad \vdots & \\
S_{n-1}(x), & x_{n-1} \leq x \leq x_n
\end{cases}
\]
Each segment is a cubic polynomial:
\[
S_j(x) = a_j + b_j (x - x_j) + c_j (x - x_j)^2 + d_j (x - x_j)^3, \quad j = 1, \ldots, n-1
\]

The coefficients \( a_j, b_j, c_j, d_j \) are determined by solving a linear system derived from:
\begin{itemize}
  \item Function values at the data points: \( f(x_i) = y_i \)
  \item Continuity of the function and its first and second derivatives at each interior knot:
  \[
  S_{j-1}(x_j) = S_j(x_j), \quad S_{j-1}'(x_j) = S_j'(x_j), \quad S_{j-1}''(x_j) = S_j''(x_j)
  \]
  \item Natural boundary conditions: \( f''(x_1) = 0 \), \( f''(x_n) = 0 \)
\end{itemize}

Once the coefficients are obtained, the spline and its derivatives on interval \( [x_j, x_{j+1}] \) are:
\[
f'(x) = b_j + 2c_j(x - x_j) + 3d_j(x - x_j)^2
\]
\[
f''(x) = 2c_j + 6d_j(x - x_j)
\]

This procedure fully defines the function computed by \texttt{smooth.spline} in R.




\subsection{Step-by-step R Code}

\vspace{1em}
\textbf{Step 1: Obtain the Spatial Field from the Model (Function \texttt{get\_field})}

\begin{verbatim}
get_field <- function(range.fraction, id.node, prior.range) {
tbm <- inla.tbm.pcmatern(
mesh = mesh,
fem = fem,
barrier.triangles = barrier.triangles,
prior.range = prior.range,
prior.sigma = prior.sigma,
range.fraction = range.fraction
)

Q <- inla.rgeneric.q(
tbm, "Q",
theta = c(log(prior.sigma[1]), log(prior.range[1]))
)

sd <- sqrt(diag(inla.qinv(Q)))
Inode <- rep(0, dim(Q)[1])
Inode[id.node] <- 1

covar.column <- solve(Q, Inode)
field <- drop(matrix(covar.column)) / (sd * sd[id.node])

return(field)
}
\end{verbatim}

In this case \texttt{range.fraction = 1} as we are using the prior range of the normal area as the range for the entire study region. \texttt{id.node} is the mesh node we are using as the reference spatial point to calculate correlation.

\vspace{1em}
\textbf{Step 2: Project the Field and Calculate Correlation with Distance (Function \texttt{get\_correlation\_df})}

\begin{verbatim}
get_correlation_df <- function(field, dims = 300, id.coord) {
proj <- inla.mesh.projector(
mesh, xlim = xlim, ylim = ylim, dims = c(dims, dims)
)

field.proj <- inla.mesh.project(proj, field)
corr <- as.vector(field.proj)
locs <- proj$lattice$loc
dist <- sqrt((locs[,1] - id.coord[[1]])^2 + (locs[,2] - id.coord[[2]])^2)

df <- data.frame(x = locs[,1], y = locs[,2], dist = dist, corr = corr)
return(df)
}
\end{verbatim}

Where \texttt{id.coord} are the coordinates of the mesh node \texttt{id.node}.

\vspace{1em}
\textbf{Step 3: Fit a Smoothing Spline to Correlation Data from Normal Area}

Taking into account \texttt{df} obtained in step 3, \texttt{s} the scaling factor, \texttt{$c_0=0.5$} the reference correlation, and \texttt{$t_{c_0}=0.2$} the desired transparency at \texttt{$c_0$}, we can build the following example:

\begin{verbatim}
ratio_fun <- function(df, s, c0, t0) {
  # Fit original spline
  spline_fit <- smooth.spline(df$field, df$dist)
  
  # Fit spline with faster decay (scaled spline)
  spline_fit_fast <- smooth.spline(df$field / s, df$dist)
  
  # Get predictions at c0
  y_at_c0 <- predict(spline_fit, x = c0)$y
  y_at_c0_fast <- predict(spline_fit_fast, x = c0)$y
  
  # Compute ratio
  ratio <- y_at_c0_fast / y_at_c0
  return(ratio - t0)
}
\end{verbatim}

Then, we find \texttt{s} such that the ratio at \texttt{$c_0$} is t0

\begin{verbatim}
c0 <- 0.5
t0 <- 0.2

result <- uniroot(function(s) ratio_fun(df, s, c0, t0), interval = c(1.01, 20))
s_solution <- result$root
cat("Scaling factor s needed:", s_solution, "\n")
\end{verbatim}

From the code we get \texttt{Scaling factor s needed: 1.864426}. 

Next we check the results as follows

\begin{verbatim}
#Fit scaled spline and check the result
spline_fit <- smooth.spline(df$field, df$dist)
spline_fit_scaled <- smooth.spline(df$field / s_solution, df$dist)
y_at_c0 <- predict(spline_fit, x = c0)$y
y_at_c0_fast <- predict(spline_fit_scaled, x = c0)$y
check_s <- y_at_c0_fast / y_at_c0
cat("Final ratio (should be", t0, "):", check_s, "\n")
\end{verbatim}

From the code we get \texttt{Final ratio (should be 0.2 ): 0.2000016}.

Lastly we use the scaled spline to find the distance at which correlation is 0.13, and use it to find the range fraction $p_b$

\begin{verbatim}
c0.13 <- 0.13
y_at_0.13 <- predict(spline_fit, x = c0.13)$y
y_at_0.13_scaled <- predict(spline_fit_scaled, x = c0.13)$y
rb_empirical <- y_at_0.13_scaled / y_at_0.13
cat("Range fraction at t0 =", t0, "and c0 =", c0, "is",  rb_empirical, "\n")
\end{verbatim}

From the code we get \texttt{Range fraction at t0 = 0.2, and c0 = 0.5 is 0.7475286}, and we use it to fill Table~\ref{tab1}.

\section{Appendix B. Sensitivity Analysis for Prior Specifications}\label{secB1}

Sensitivity analysis is a crucial step in Bayesian statistical modeling, particularly when assessing the impact of prior distributions on posterior inference. In spatial modeling, the choice of priors can significantly influence the estimates of parameters, potentially affecting the model's interpretation and conclusions \cite{berger_objective_2001, ferreira_bayesian_2007, simpson_penalising_2017, fuglstad_constructing_2019}. Sensitivity analysis helps to understand the robustness of model results to different prior assumptions and provides insights into how changes in prior information can affect posterior distributions \cite{greenland_sensitivity_2001, oakley_probabilistic_2004}.

In this appendix, we present a sensitivity analysis conducted on the spatial parameters of our Transparent Barrier model (SGFs) applied to dugong species distribution in the Red Sea. We specifically focus on the impact of varying barrier permeability and different prior choices for the range parameter. 

Table~\ref{tab:sensi1}, Table~\ref{tab:sensi2}, Table~\ref{tab:sensi3}, Table~\ref{tab:sensi4} summarize posterior estimates for the 
spatial range parameter and standard deviation $\sigma$ under varying barrier permeability conditions ranging from fully permeable to increasingly impermeable barriers. Eight distinct priors for the spatial range parameter have been tested to assess their influence on the posterior estimates. These priors reflect various levels of certainty regarding the spatial correlation, ranging from strong beliefs in short-range spatial dependence to assumptions of long-range spatial influence.

The results demonstrate that as barriers become more impermeable, posterior estimates consistently reflect reduced spatial correlation across barriers, highlighting the effectiveness of the barrier model in adjusting spatial correlation according to physical barriers. Variations in priors have a greater effect on the posterior distribution as barrier impermeability increases; with higher barrier impermeability, priors indicating stronger short-range spatial correlation yield narrower credible intervals, demonstrating increased confidence in short-range spatial dependence. Under highly permeable barrier conditions, posterior distributions for the range parameter exhibit relatively minimal sensitivity to prior assumptions, indicating robustness of the model in scenarios where spatial connectivity is less restricted.

The sensitivity analysis highlights that both the condition of barrier permeability and the selection of prior distributions substantially influence posterior inference in Bayesian spatial modeling. Our analysis indicates that posterior estimates are most sensitive to prior specifications when barriers exhibit strong impermeability. This is due to the very low number of data points available in those regions, which makes the posterior more dependent on prior assumptions. Consequently, careful consideration of barrier conditions and informed prior choice are essential for accurate spatial modeling, especially when working with physically constrained regions where observational data are sparse.

\begin{table}[htbp]
\centering
\caption{Posterior estimates under fully permeable barriers conditions (barrier 1 = 1, barrier 2 = 1, barrier 3 = 1)}
\label{tab:sensi1}
\begin{tabular*}{\textwidth}{@{\extracolsep\fill}clrrrrr}
Parameter & $u $, $\sigma_{u }$ & Mean & Mode & 0.5 Quant & 0.025 Quant & 0.975 Quant \\
\cmidrule[0.1pt]{1-7}
$\sigma$  & 1, 0.1   & 1.23 & 1.21 & 1.23 & 0.88 & 1.77 \\
range     & 21, 0.5  & 7.56 & 7.63 & 7.60 & 4.32 & 12.80 \\
\cmidrule[0.1pt]{1-7}
$\sigma$  & 1, 0.1   & 1.37 & 1.36 & 1.37 & 1.13 & 1.67 \\
range     & 21, 0.9  & 6.08 & 6.00 & 6.06 & 4.59 & 8.17 \\
\cmidrule[0.1pt]{1-7}
$\sigma$  & 1, 0.1   & 1.13 & 1.12 & 1.13 & 0.85 & 1.52 \\
range     & 41, 0.5  & 9.10 & 9.44 & 9.17 & 4.99 & 15.95 \\
\cmidrule[0.1pt]{1-7}
$\sigma$  & 1, 0.1   & 1.32 & 1.33 & 1.32 & 1.07 & 1.63 \\
range     & 41, 0.9  & 6.68 & 6.34 & 6.63 & 4.61 & 10.29 \\
\cmidrule[0.1pt]{1-7}
$\sigma$  & 1, 0.1   & 1.32 & 1.30 & 1.33 & 1.06 & 1.68 \\
range     & 10, 0.5  & 6.19 & 6.53 & 6.23 & 3.32 & 10.82 \\
\cmidrule[0.1pt]{1-7}
$\sigma$  & 1, 0.1   & 1.39 & 1.38 & 1.39 & 1.16 & 1.69 \\
range     & 10, 0.9  & 5.85 & 5.78 & 5.83 & 4.43 & 7.83 \\
\cmidrule[0.1pt]{1-7}
$\sigma$  & 1, 0.1   & 1.34 & 1.34 & 1.34 & 1.05 & 1.72 \\
range     & 6, 0.5   & 6.35 & 6.24 & 6.33 & 3.15 & 13.04 \\
\cmidrule[0.1pt]{1-7}
$\sigma$  & 1, 0.1   & 1.38 & 1.37 & 1.38 & 1.00 & 1.92 \\
range     & 6, 0.9   & 5.72 & 5.77 & 5.73 & 3.43 & 9.45 \\
\bottomrule
\end{tabular*}
\end{table}


\begin{table}[!htbp]
\centering
\caption{Posterior estimates with barrier 1 highly impermeable, others fully permeable (barrier 1 = 0.01, barrier 2 = 1, barrier 3 = 1)}
\label{tab:sensi2}
\begin{tabular*}{\textwidth}{@{\extracolsep\fill}clrrrrr}
Parameter & $u $, $\sigma_{u }$ & Mean & Mode & 0.5 Quant & 0.025 Quant & 0.975 Quant \\
\midrule
\(\sigma\)& 1, 0.1 & 1.16 & 1.14 & 1.15 & 0.69 & 1.95 \\
range     & 21, 0.5  & 49.49 & 37.92 & 47.91 & 16.32 & 201.98 \\
\midrule
\(\sigma\)& 1, 0.1 & 1.15 & 1.19 & 1.16 & 0.68 & 1.90 \\
range     & 21, 0.9  & 38.54 & 31.38 & 37.17 & 13.68 & 135.34 \\
\midrule
\(\sigma\)& 1, 0.1 & 1.12 & 1.12 & 1.12 & 0.63 & 2.00 \\
range     & 41, 0.5  & 52.68 & 43.98 & 50.82 & 16.13 & 208.30 \\
\midrule
\(\sigma\)& 1, 0.1 & 1.15 & 1.16 & 1.15 & 0.67 & 1.95 \\
range     & 41, 0.9  & 37.81 & 34.08 & 37.03 & 13.68 & 116.83 \\
\midrule
\(\sigma\)& 1, 0.1 & 1.17 & 1.15 & 1.16 & 0.70 & 1.95 \\
range     & 10, 0.5  & 44.67 & 35.22 & 43.21 & 15.39 & 169.08 \\
\midrule
\(\sigma\)& 1, 0.1 & 1.17 & 1.18 & 1.17 & 0.70 & 1.90 \\
range     & 10, 0.9  & 37.81 & 31.09 & 36.32 & 12.71 & 137.68 \\
\midrule
\(\sigma\)& 1, 0.1  & 1.15 & 1.17 & 1.16 & 0.68 & 1.93 \\
range     & 6, 0.5  & 37.24 & 33.02 & 36.31 & 13.27 & 118.99 \\
\midrule
\(\sigma\)& 1, 0.1  & 1.17 & 1.17 & 1.17 & 0.71 & 1.93 \\
range     & 6, 0.9  & 39.99 & 32.55 & 38.46 & 14.10 & 147.10 \\
\bottomrule
\end{tabular*}
\end{table}

\begin{table}[!htbp]
\centering
\caption{Posterior estimates with barrier 1 highly impermeable, barrier 2 moderately permeable, and barrier 3 fully permeable (barrier 1 = 0.01, barrier 2 = 0.1, barrier 3 = 1).}
\label{tab:sensi3}
\begin{tabular*}{\textwidth}{@{\extracolsep\fill}clrrrrr}
Parameter & $u $, $\sigma_{u }$ & Mean & Mode & 0.5 Quant & 0.025 Quant & 0.975 Quant \\
\midrule
\(\sigma\) & 1, 0.1 & 1.01 & 1.07 & 1.02 & 0.67 & 1.45 \\
range      & 21, 0.5  & 42.07 & 28.08 & 40.22 & 15.01 & 184.02 \\
\midrule
\(\sigma\) & 1, 0.1 & 1.09 & 1.08 & 1.09 & 0.71 & 1.72 \\
range      & 21, 0.9  & 23.93 & 24.76 & 24.09 & 11.03 & 50.06 \\
\midrule
\(\sigma\) & 1, 0.1 & 1.06 & 1.06 & 1.06 & 0.66 & 1.69 \\
range      & 41, 0.5  & 36.03 & 30.29 & 34.82 & 14.65 & 106.18 \\
\midrule
\(\sigma\) & 1, 0.1 & 1.09 & 1.09 & 1.09 & 0.69 & 1.69 \\
range      & 41, 0.9  & 27.30 & 24.93 & 26.80 & 12.82 & 64.14 \\
\midrule
\(\sigma\) & 1, 0.1 & 1.08 & 1.09 & 1.08 & 0.69 & 1.69 \\
range      & 10, 0.5  & 28.10 & 25.43 & 27.57 & 13.01 & 67.65 \\
\midrule
\(\sigma\) & 1, 0.1 & 1.09 & 1.09 & 1.09 & 0.71 & 1.68 \\
range      & 10, 0.9  & 27.13 & 24.21 & 26.61 & 12.73 & 65.47 \\
\midrule
\(\sigma\) & 1, 0.1  & 1.10 & 1.08 & 1.09 & 0.69 & 1.76 \\
range      & 6, 0.5  & 24.73 & 24.40 & 24.66 & 12.10 & 81.28 \\
\midrule
\(\sigma\) & 1, 0.1  & 1.08 & 1.09 & 1.09 & 0.71 & 1.65 \\
range      & 6, 0.9  & 30.63 & 24.60 & 29.80 & 12.84 & 93.27 \\
\bottomrule
\end{tabular*}
\end{table}

\begin{table}[!htbp]
\centering
\caption{Posterior estimates with all barriers exhibiting impermeability. Barrier 1 is fully impermeable, and barrier 2  and 3 with increasing permeability (barrier 1 = 0.01, barrier 2 = 0.1, barrier 3 = 0.2).}
\label{tab:sensi4}
\begin{tabular*}{\textwidth}{@{\extracolsep\fill}clrrrrr}
Parameter & $u $, $\sigma_{u }$ & Mean & Mode & 0.5 Quant & 0.025 Quant & 0.975 Quant \\
\midrule
\(\sigma\) & (1, 0.1)   & 1.14 & 1.13 & 1.14 & 0.64 & 2.08 \\
range      & (21, 0.5)  & 44.66 & 39.70 & 43.58 & 14.22 & 159.20 \\
\midrule
\(\sigma\) & (1, 0.1)   & 1.14 & 1.15 & 1.15 & 0.66 & 1.96 \\
range      & (21, 0.9)  & 38.91 & 34.94 & 38.09 & 13.73 & 123.85 \\
\midrule
\(\sigma\) & (1, 0.1)   & 1.11 & 1.12 & 1.11 & 0.61 & 1.98 \\
range      & (41, 0.5)  & 54.26 & 47.36 & 52.90 & 17.19 & 198.63 \\
\midrule
\(\sigma\) & (1, 0.1)   & 1.09 & 1.14 & 1.10 & 0.72 & 1.56 \\
range      & (41, 0.9)  & 43.19 & 35.62 & 42.98 & 15.39 & 152.77 \\
\midrule
\(\sigma\) & (1, 0.1)   & 1.14 & 1.15 & 1.14 & 0.66 & 1.96 \\
range      & (10, 0.5)  & 41.39 & 36.61 & 40.38 & 14.08 & 138.87 \\
\midrule
\(\sigma\) & (1, 0.1)   & 1.16 & 1.17 & 1.16 & 0.70 & 1.91 \\
range      & (10, 0.9)  & 43.79 & 33.27 & 42.74 & 14.96 & 174.49 \\
\midrule
\(\sigma\) & (1, 0.1)   & 1.15 & 1.15 & 1.15 & 0.68 & 1.96 \\
range      & (6, 0.5)   & 44.04 & 35.57 & 42.52 & 14.83 & 165.36 \\
\midrule
\(\sigma\) & (1, 0.1)   & 1.16 & 1.15 & 1.15 & 0.66 & 2.05 \\
range      & (6, 0.9)   & 37.82 & 33.83 & 36.95 & 13.15 & 122.69 \\
\bottomrule
\end{tabular*}
\end{table}

\newpage

\bibliographystyle{unsrt}  
\bibliography{references1}



\end{document}